\documentclass[]{aa}         

\usepackage[final]{epsfig} 
\usepackage{times}
\usepackage{graphics}

\usepackage{natbib}
\bibpunct{(}{)}{;}{a}{}{,}

\usepackage[american,german,english]{babel}

\usepackage{latexsym}
\usepackage{amssymb}

\begin{document}

\newcommand{\iras}{\object{IRAS~04381+2540}}

\title{HST/NICMOS observations of a proto-brown dwarf candidate}
 \author {
         D\'aniel Apai\inst{1,2,3}
        \and L. Viktor T\'oth\inst{1,4,5}
        \and Thomas Henning\inst{1}
        \and Roland Vavrek\inst{6} 
        \and Zolt\'an Kov\'acs\inst{1}
        \and Dietrich Lemke\inst{1}
            } 
     \institute{Max Planck Institute for Astronomy,
                K\"onigstuhl 17,
                D-69117, Heidelberg, Germany
                \and
                Steward Observatory, The University of Arizona, 933 N. Cherry Avenue, Tucson, AZ 85721, USA     
                \and
                NASA Astrobiology Institute
		\and
                Department of Astronomy of the Lor\'and E\"otv\"os
                University, P\'azm\'any P\'eter s\'et\'any 1, 
                H-1117 Budapest,  Hungary
                \and
                Konkoly Observatory of the Hungarian Academy of Sciences, PO Box 67, 1525 Budapest, Hungary
		\and
                Astrophysics Missions Division, ESTEC, Keplerlann 1, Norwijk, NL-2201, The Netherlands
                }
     \offprints{D.~Apai~{apai@as.arizona.edu}}

\titlerunning{}
     

\abstract{
We present deep HST/NICMOS observations  peering through the outflow cavity 
of the protostellar candidate IRAS~04381+2540 in the Taurus Molecular Cloud-1. A young stellar object as central source, 
a jet and a very faint and close (0.6") companion are identified.
The primary and the companion have similar colours, consistent
with strong reddening. We argue that the companion is neither a shock-excited knot nor a background star.
The colour/magnitude information predicts a substellar upper mass limit for the companion, but the final confirmation will require spectroscopic information.  Because of its geometry, young age and its rare low-mass companion,
this system is likely to provide a unique insight into the formation of brown dwarfs.
}%

 \maketitle

\section{Introduction}

The debate on the origin of brown dwarfs has preceded their discovery \citep{1995Natur.377..129R}: their small masses poses serious difficulty to a stellar-like formation scenario through the collapse of a molecular cloud core.  Now,  with hundreds of known brown dwarfs, the debate is still not settled: the most widely discussed formation scenarios include gravitational collapse in highly turbulent cloud cores (e.g. \citealt{2004ApJ...617..559P}), early ejection from unstable multiple systems \citep{1998A&A...339...95S,2001AJ....122..432R,Umbreit} and photoevaporation of protostars \citep{2004A&A...427..299W}.
 Only few traces of the formation process survive the early evolution of these objects. Current studies
focused on disk indicators, signs of accretion, binarity and velocity dispersion. 
We learned that brown dwarfs are often surrounded by warm circumstellar material (e.g. \citealt{2003AJ....126.1515J}), with masses of $\sim$M$_{\mathrm Jup}$ \citep{2003ApJ...593L..57K} which they maintain over timescales similar to stars \citep{2004A&A...427..245S}, and in one case 
it could be shown that the dust forms a disk \citep{2003ApJ...590L.111P}. 
Because gravitational ejection leaves sufficient material in the inner disk \citep{2002MNRAS.332L..65B}  to produce infrared excess emission and accretion signs, the ejection embryo  hypothesis remains a possible formation scenario.
The masses derived from millimetre-wavelength observations \citep{2003ApJ...593L..57K} are too uncertain to constrain possible disk truncation and therefore cannot exclude the  ejection scenario.

Direct observations of the youngest phases of substellar objects, termed as proto-brown dwarfs, will be crucial in evaluating the formation scenarios. Recent  imaging by \citet{2004AJ....127.1747H} and \citet{2004A&A...427..651D} identified faint, possibly substellar companion candidates to protostellar systems, 
but some of them might be background stars. To ensure the very young age of the proto-brown dwarfs, they should be observed
within the massive protostellar envelope, possible only in exceptional cases.

We present here observations of the first such system, \iras,
 located in \object{TMC-1} in the Taurus star-forming region with the  discovery of a proto-brown dwarf candidate.
Based on its spectral energy distribution and bolometric temperature,
\iras{} has been classified as a Class\,I object \citep{2000ApJ...534..880H}. However,
estimates of the stellar mass ($M=0.3\pm 0.1$M$_{\odot }$, \citealt{1999MNRAS.303..855B})
show that it is comparable to the envelope mass ($M=0.18$M$_{\sun }$, \citealt{2003ApJS..145..111Y})
 indicating that the object
is still in its main accretion phase \citep{2001A&A...365..440M}.
\iras{} drives a molecular outflow \citep{1999MNRAS.303..855B},  with an inclination of $i\approx 50\degr$ 
 (the northern part tilted toward the observer, \citealt{1996ApJ...471..308C}).
 Through the evacuated outflow cone we see into the dense molecular cloud core   
in which \iras{} is embedded, making it an  exceptional target 
for studies of very early stellar evolution.

\section{Observations and data reduction}
\label{Observations}
We used archival HST/NICMOS observations to study \iras{}  at
small spatial scales. The main parameters of the observations are summarised
in Tab.~\ref{log}.

\begin{table*}
\begin{center}
\begin{tabular}{llcccrrrrl}
\hline
\hline
Instrument&Filter&Date    & Field of    & Exp. Time  &  \multicolumn{2}{c}{Flux of NIR~A}&\multicolumn{2}{c}{Flux of NIR~B}& Note  \\
          &      &DD/MM/YY        & View    &[s]     &     [$\mu$Jy]& [mag]& [$\mu$Jy]   &[mag]&  \\
 \hline
HST/NICMOS& F160W&20/12/97&20\arcmin$\times$19\arcmin& 1280& 511  & 15.8 &38  &18.6 &1.4-1.8$\mu $m \\
HST/NICMOS& F205W&07/01/98&19\arcmin$\times$19\arcmin& 256 &6295 & 12.6 &363  &15.7&1.75-2.35$\mu $m \\
HST/NICMOS& F212N&07/01/98&19\arcmin$\times$19\arcmin& 160 & 7711 & 12.3 & $<$1070  & $>$14.5&2.121$\mu $m,~H$_2$\\
HST/NICMOS& F212N&07/01/98&19\arcmin$\times$19\arcmin& 160 &      &      & $\leq$351  &  & 0.13" aper. \\
\hline

\end{tabular}
\caption[]{Log of the infrared observations and results of the NICMOS photometry. The estimated photometric error
is less than 10\% for all bands. In the F212N band only an upper limit can be reliably established due to the low 
contrast between the object and the background nebulosity. Here, the use of a smaller aperture reduces the
contamination from the nebulosity and places a stricter upper limit. 
The F160W filter also includes [Fe~II] line emission at 1.644~$\mu$m.\label{log} }
\end{center}
\end{table*}
The data have been successfully pipeline-processed and required only minor
additional reduction work, which was carried out by simple IDL scripts.
The images have been bad pixel filtered and the central columns  
bias-corrected in the standard fashion.
The final mosaic has been composed by cross-correlating the individual
pointings. 
In order to enhance the fainter details around the central point source
we subtracted synthetic point spread functions (PSF) calculated by using the
Tiny Tim 6.1 \citep{1997hstc.work..192K}.  The
scaling factors have been obtained by comparing the integrated counts in 
3-pixel-radii apertures placed on the simulated and observed peaks.

Aperture photometry has been performed on all images in an identical way. 
We used the IDL adaptation of the DAOPHOT routine
to integrate the counts in 6-pixel-radius apertures centered on the
central point source. 
In order to obtain reliable photometry also for the fainter, 
secondary point source we carried out PSF photometry by subtracting
simulated PSFs. By minimizing the 
subtraction residuals we measured the point source's brightness without significant 
contribution
from the surrounding nebulosity. In the case of the F212N filter, due to the faintness of the secondary point source compared to the nebula,
we could only estimate upper limits from the aperture photometry. Here, we also applied a second, small 3$\times$3 pixel aperture
to suppress the contamination from the nebula and thus obtained a stricter upper limit. 

\begin{figure}
\epsfig{file=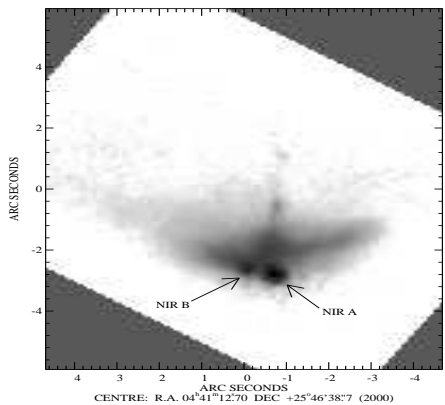,height=10.0cm,width=6cm,angle=90}
\caption{High-resolution near-infrared view of \iras{} by HST/NICMOS in the broadband F160W filter.
This image reveals a binary young stellar object (NIR~A+B) and a collimated  jet.
\label{NICMOSH}}
\end{figure}
\section{Results}

As shown in Fig.~\ref{NICMOSH} the  NICMOS image reveals a detailed view of the reflection
nebulosity including two point sources (\object{IRAS 04381+2540~NIR~A} and \object{IRAS 04381+2540~NIR~B}) 
and a long optical jet at the heart of IRAS 04381+2540. In the following, we will refer to the brighter, primary point 
source as  NIR~A and to the fainter, secondary one as NIR~B. 
NIR~A is located in the apparent center of symmetry of the reflection nebula
with the position of RA: 04$^h$ 41$^m$ 12\farcs65 DEC: +25$\degr$ 46$'$ 35\farcs93 (J2000). 
This position is accurate within the spacecraft's pointing accuracy, typically better than 0\farcs5.
All NICMOS images show NIR~A  as a point source at a resolution of 0\farcs15.
The secondary source NIR~B is located at 0\farcs59 east and 0\farcs15 north 
of NIR~A. The source is unresolved in all three NICMOS images, 
while in the F160W image a possible marginal, faint extension is seen at the brightness level of the surrounding nebula, possibly a faint scattered-light envelope of NIR~B. The photometry results are summarized in Table~\ref{log}.
The deepest F160W image also reveals the presence of a $\sim$3\farcs3 long, narrow line of intense
emission, which we identified as 
a highly collimated jet. While the northern end of the jet vanishes in the background
noise about 5\arcsec{}  from NIR~A, its southern end is overshined by the reflection nebula at
about 1.2\arcsec{} from NIR~A. The jet's direction points to NIR~A identifying its driving source.

\section{Discussion}

 The detection of the collimated jet proves that NIR~A, its driving source, is still
 in the early, accreting phase of its evolution.  By assuming an inclination
 of $i=50\degr$ for the jet its visible length is $\sim 1000~$AU.
  As often observed, the optical jet in
 IRAS~04381+2540 is accompanied by a molecular outflow \citep{1996ApJ...471..308C}. 
The present and former studies suggest the following picture: the young stellar object
 NIR~A -- at a transition stage in the Class 0/Class I boundary -- is located close to the side of the dense molecular cloud core facing the observer.
  The low-mass source NIR~A drives a collimated jet and harbors a circumstellar accretion disk
  perpendicular to the jet's direction.  The jet-powered outflow clears out a low-density cavity,
  whose wall becomes bright by scattering the light of NIR~A.
The southern jet is inclined away from the observer and, by penetrating into the denser regions of the cloud, 
 becomes strongly extincted and apparently fainter.
  The northern beam of the jet is inclined toward the observer and is visible inside the outflow cavity.
The  age of the system can be estimated on the basis of the $\sim10^5$~yr typical lifetime of protostellar 
objects, consistent e.g. with the chemical evolution studies of the deuterium fractionation \citep{2002ApJ...569..836S} in \iras{}.

\begin{figure}
\epsfig{file=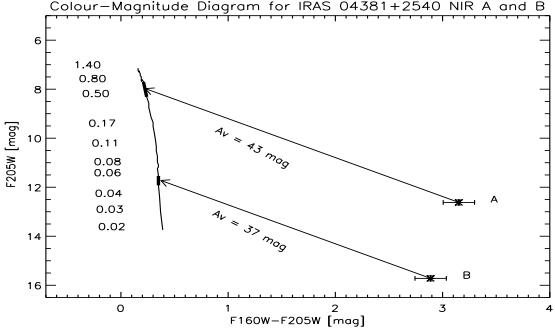, width=8cm,height=4.5cm}
\caption{Colour-magnitude diagram based on HST/NICMOS photometry 
for the IRAS~04381+2540~ NIR~A and B sources. The 1~Myr-old 
isochrone from \citet{1998A&A...337..403B} is overplotted. The numbers indicate the masses
in solar units.  The  interstellar 
reddening vectors for A$_{\mathrm V}$=43~mag and 37~mag are also plotted.
Thick lines mark the approximate mass range as constrained by the photometric errors.
  \label{colmag}}
\end{figure}

\subsection{The nature of the point sources}
 The near-infrared colour information allows preliminary conclusions
on the nature of NIR~A. When plotted in a colour-magnitude diagram (see Fig.~\ref{colmag}) 
the photometry of NIR~A is consistent with that of a highly reddened (A$_V\approx$ 43~mag) 
young stellar object. The mass estimate of $\sim$0.6~M$_\odot$ 
based on the comparison of the near-infrared photometry
to the 1~Myr isochrones of \citet{1998A&A...337..403B} and is consistent with 
the observed bolometric luminosity (0.7 L$_\odot$, \citealt{1998MNRAS.299..789C}). A
similar mass estimate was derived from gas kinematics (0.3$\pm$0.1 M$_\odot$, \citealt{1999MNRAS.303..855B}).
The case of NIR~B is  more compelling; as NIR~B does not have any 
evident link to its environment suggesting it to be a companion of NIR~A, 
it could also be an unrelated background/foreground star, or an excited knot in the
cavity wall. These three possibilities are briefly discussed below.

\noindent{(1) \em Foreground/background source}: The high reddening 
of NIR~B (A$_V\approx$37~mag)  excludes the possibility 
of an unrelated foreground star. The comparison \citep{2004A&A...425..133B} of the \citet{1992ApJS...83..111W}  
model to the 2MASS~Point Source Catalogue  predicts
$A_{\mathrm V}\approx$16~mag extinction towards our source and  
the probability of a chance background star appearing within $\sim$1.5 square arcsecond to be $\approx 10^{-4}$.
The additional A$_{\mathrm V}\gtrsim$20~mag extinction observed towards \iras{} makes such a 
coincidence even less likely.

\noindent{(2) \em A shock-excited knot associated to the jet:} 
 Although several such examples are known, e.g., in L1551~NE (e.g. \citealt{2000AJ....120.1449R}),
 the source NIR~B differs from these objects in three important aspects: it is visible
 and unresolved in the broad-band images; and its brightness in the F212N filter is dominated by 
continuum emission (i.e. no shocked H$_2$ emission is visible).

\noindent{(3) \em Companion to NIR~A:}  NIR~B has near-infrared colours very similar to NIR~A; a comparison
to the 1~Myr-old substellar models of \citet{1998A&A...337..403B} suggests a reddening almost identical
to NIR~A (see Fig.~\ref{colmag}). In addition to the matching colours, the very proximity of NIR A and B further
reinforces their physical link.
The alternative scenarios being very improbable, in the following we tentatively 
accept NIR~B to be a point source related to NIR~A. 

\subsection{Probable substellar mass for NIR~B}
The comparison of NIR~B brightness and colour (see Fig.~\ref{colmag}) argues for a substellar
mass of M$_{B}\simeq 50\pm 4$~M$_{\mathrm{Jup}}$. The error bar represents the photometric error, but in the following
we discuss  several other factors which influence this estimate.
 
\noindent{\em (1) Isochrones:} Our object is likely to be younger than the youngest available isochrones dating
1~Myr. The yet unknown initial conditions of brown dwarf formation probably introduce
significant uncertainities even for the 1~Myr tracks \citep{2002A&A...382..563B}. 
Comparing the luminosity and effective temperature in the 40--60~M$_{{\mathrm Jup}}$ range to the 1~Myr-old models by 
\citet{1997ApJ...491..856B}, we find that the latter models associate 3--5~M$_{{\mathrm Jup}}$ smaller masses to the given luminosities, suggesting an even smaller mass for NIR~B. 
Although no isochrones of the appropriate age are yet available, using overaged
isochrones lead to  the {\em overestimation} of the mass and therefore enables us to set an 
upper mass limit for NIR~B of $\simeq$50~M$_{{\mathrm Jup}}$.

\noindent{\em (2) Reddening:} The reddening law plays a crucial role when interpreting 
objects as heavily reddened as NIR~B. In Fig.~\ref{colmag} we used the reddening law by
\citet{1990ARA&A..28...37M} and now we explore how the estimate varies with the reddening law's
slope. A study of the reddening toward the Taurus region by 
\citet{2001ApJ...547..872W} showed that in the direction of dense clouds R$_{\mathrm V}$= 3.5 -- 4.0 and might reach 4.5
 for the densest regions of the cloud. Disentangling the
contribution to NIR~B's reddening by the different dust types is not possible. By making the worst-case
estimate (assuming that all the extinction is caused by dust with R$_{\mathrm V}$ as high as 4.5)
would modify NIR~B's upper mass limit to $\simeq$90~M$_{{\mathrm  Jup}}$, i.e. slightly above the substellar limit.
We conclude, that although the reddening law influences the mass estimates, it is very unlikely to shift
the upper mass limit into the stellar regime.

\noindent {\em (3) Anisotropic emission:} The emission of embedded stars is often influenced by the
circumstellar environment via obscuration and light scattering, exemplified by edge-on
disks (see, e.g. \citealt{1999AJ....117.1490P}). In these cases both the colour and the magnitude of the 
objects are  altering mass estimates on the basis of theoretical isochrones.
Circumstellar material obscuring more than 66\% of NIR~B's emission could  apparently shift  the luminosity of a very low mass star to substellar luminosities. Alternatively, NIR~B's emission might originate partly by its envelope,  scattering the light of NIR~A; then
NIR~B's mass might be even smaller than estimated. Entirely excluding these possibilities is impossible without
detailed spectroscopic and polarimetric follow-up observations but we argue that such a strong scattered-light nebula would likely be resolvable by our observations.  While a study by \citet{1999AJ....117.1490P} shows 7 such nebulae around edge-on disks 
 to have typical sizes of $\simeq$800~AU, NIR~B is an unresolved point source at our $\sim$20~AU resolution.
Thus, should NIR~B be a reflection nebula, its point source nature would be rather exceptional. 

\subsection{NIR~B as a brown dwarf in formation} 
The concept of NIR~B being a substellar companion of NIR~A opens up exciting
possibilities. Although several stars with brown dwarf companions are known, this object  
is about an order of magnitude younger than the previously known youngest brown dwarf companion to the 
T Tauri star GG Tau Bb \citep{1999ApJ...520..811W}. The presence of the optical jet confirms active
accretion in the system and suggests that NIR~B is a brown dwarf in formation. 
An important question
is whether NIR~B will be able to accrete sufficient mass to become a star during the future
accretion processes. Observations of the Br$\gamma$ line can determine the accretion rates of NIR~A and NIR~B, and show
whether the current mass difference will increase (to produce a low-mass star with brown dwarf companion) or decrease (toward an equal-mass binary situation). 
Additional, longer wavelength high-resolution observations can identify whether NIR~B is embedded in the disk of NIR~A or harbors its own accretion disk, an important clue on the formation history of the system. Should NIR~B be in the disk, it is likely to gravitationally interact with the disk
and its mass is probably sufficient to induce observable structures,
similar to the ones predicted for massive planets embedded in Solar-type
stars' disks (e.g. \citealt{2000ApJ...540.1091B,2003ApJ...586..540D}) allowing the 
observational test of these models.

\section{Conclusions}
Based on our HST/NICMOS infrared imaging the main conclusions of this work
are the following:

\noindent (i) The central source of the IRAS 04381+2540 system is identified as a young, 
low-mass binary stellar object.

\noindent (ii) The accreting primary component drives a collimated jet.

\noindent (iii) The secondary component is probably the youngest known brown dwarf so far, providing a unique
opportunity to study the formation of substellar objects.

\begin{acknowledgements}
\sloppy
{\small 
The HST data have been obtained in the frame of the
programmes 7325 and 7413 (PIs: L. Hartmann and S. Terebey). We are grateful to I. Baraffe for providing the isochrone 
data for the HST filter set. Valuable comments by I.~Pascucci, W. Brandner, G. Schneider, L. Close, J. Muzerolle, R.~Mundt and M. Goto, and technical help by 
T. Khazadyan are acknowledged. The prompt and constructive review by our referee, G. Duch\^ene strengthened the science
case and improved its presentation.
This material is partly based upon work supported by NASA through the NASA Astrobiology Institute under
Cooperative Agreement No. CAN-02-OSS-02. This research was partly supported by the
OTKA grant T-043774.}
\end{acknowledgements}

\bibliographystyle{aa}
\bibliography{lit.bib}

\begin{thebibliography}{33}
\expandafter\ifx\csname natexlab\endcsname\relax\def\natexlab#1{#1}\fi

\bibitem[{{Bal{\' a}zs} {et~al.}(2004){Bal{\' a}zs}, {{\' A}brah{\' a}m},
  {Kun}, {Kelemen}, \& {T{\' o}th}}]{2004A&A...425..133B}
{Bal{\' a}zs}, L.~G., {{\' A}brah{\' a}m}, P., {Kun}, M., {Kelemen}, J., \&
  {T{\' o}th}, L.~V. 2004, \aap, 425, 133

\bibitem[{{Baraffe} {et~al.}(1998){Baraffe}, {Chabrier}, {Allard}, \&
  {Hauschildt}}]{1998A&A...337..403B}
{Baraffe}, I., {Chabrier}, G., {Allard}, F., \& {Hauschildt}, P.~H. 1998, \aap,
  337, 403

\bibitem[{{Baraffe} {et~al.}(2002){Baraffe}, {Chabrier}, {Allard}, \&
  {Hauschildt}}]{2002A&A...382..563B}
{Baraffe}, I., {Chabrier}, G., {Allard}, F., \& {Hauschildt}, P.~H. 2002, \aap,
  382, 563

\bibitem[{{Bate} {et~al.}(2002){Bate}, {Bonnell}, \&
  {Bromm}}]{2002MNRAS.332L..65B}
{Bate}, M.~R., {Bonnell}, I.~A., \& {Bromm}, V. 2002, \mnras, 332, L65

\bibitem[{{Brown} \& {Chandler}(1999)}]{1999MNRAS.303..855B}
{Brown}, D.~W. \& {Chandler}, C.~J. 1999, \mnras, 303, 855

\bibitem[{{Bryden} {et~al.}(2000){Bryden}, {R{\' o}{\. z}yczka}, {Lin}, \&
  {Bodenheimer}}]{2000ApJ...540.1091B}
{Bryden}, G., {R{\' o}{\. z}yczka}, M., {Lin}, D.~N.~C., \& {Bodenheimer}, P.
  2000, \apj, 540, 1091

\bibitem[{{Burrows} {et~al.}(1997){Burrows}, {Marley}, {Hubbard}, {Lunine},
  {Guillot}, {Saumon}, {Freedman}, {Sudarsky}, \&
  {Sharp}}]{1997ApJ...491..856B}
{Burrows}, A., {Marley}, M., {Hubbard}, W.~B., {et~al.} 1997, \apj, 491, 856

\bibitem[{{Chandler} {et~al.}(1998){Chandler}, {Barsony}, \&
  {Moore}}]{1998MNRAS.299..789C}
{Chandler}, C.~J., {Barsony}, M., \& {Moore}, T.~J.~T. 1998, \mnras, 299, 789

\bibitem[{{Chandler} {et~al.}(1996){Chandler}, {Terebey}, {Barsony}, {Moore},
  \& {Gautier}}]{1996ApJ...471..308C}
{Chandler}, C.~J., {Terebey}, S., {Barsony}, M., {Moore}, T.~J.~T., \&
  {Gautier}, T.~N. 1996, \apj, 471, 308

\bibitem[{{D'Angelo} {et~al.}(2003){D'Angelo}, {Kley}, \&
  {Henning}}]{2003ApJ...586..540D}
{D'Angelo}, G., {Kley}, W., \& {Henning}, T. 2003, \apj, 586, 540

\bibitem[{{Duch{\^ e}ne} {et~al.}(2004){Duch{\^ e}ne}, {Bouvier}, {Bontemps},
  {Andr{\' e}}, \& {Motte}}]{2004A&A...427..651D}
{Duch{\^ e}ne}, G., {Bouvier}, J., {Bontemps}, S., {Andr{\' e}}, P., \&
  {Motte}, F. 2004, \aap, 427, 651

\bibitem[{{Haisch} {et~al.}(2004){Haisch}, {Greene}, {Barsony}, \&
  {Stahler}}]{2004AJ....127.1747H}
{Haisch}, K.~E., {Greene}, T.~P., {Barsony}, M., \& {Stahler}, S.~W. 2004, \aj,
  127, 1747

\bibitem[{{Hogerheijde} \& {Sandell}(2000)}]{2000ApJ...534..880H}
{Hogerheijde}, M.~R. \& {Sandell}, G. 2000, \apj, 534, 880

\bibitem[{{Jayawardhana} {et~al.}(2003){Jayawardhana}, {Ardila}, {Stelzer}, \&
  {Haisch}}]{2003AJ....126.1515J}
{Jayawardhana}, R., {Ardila}, D.~R., {Stelzer}, B., \& {Haisch}, K.~E. 2003,
  \aj, 126, 1515

\bibitem[{{Klein} {et~al.}(2003){Klein}, {Apai}, {Pascucci}, {Henning}, \&
  {Waters}}]{2003ApJ...593L..57K}
{Klein}, R., {Apai}, D., {Pascucci}, I., {Henning}, T., \& {Waters},
  L.~B.~F.~M. 2003, \apjl, 593, L57

\bibitem[{{Krist} \& {Hook}(1997)}]{1997hstc.work..192K}
{Krist}, J.~E. \& {Hook}, R.~N. 1997, in The 1997 HST Calibration Workshop with
  a New Generation of Instruments, 192

\bibitem[{{Mathis}(1990)}]{1990ARA&A..28...37M}
{Mathis}, J.~S. 1990, \araa, 28, 37

\bibitem[{{Motte} \& {Andr{\' e}}(2001)}]{2001A&A...365..440M}
{Motte}, F. \& {Andr{\' e}}, P. 2001, \aap, 365, 440

\bibitem[{{Padgett} {et~al.}(1999){Padgett}, {Brandner}, {Stapelfeldt},
  {Strom}, {Terebey}, \& {Koerner}}]{1999AJ....117.1490P}
{Padgett}, D.~L., {Brandner}, W., {Stapelfeldt}, K.~R., {et~al.} 1999, \aj,
  117, 1490

\bibitem[{{Padoan} \& {Nordlund}(2004)}]{2004ApJ...617..559P}
{Padoan}, P. \& {Nordlund}, {\AA}. 2004, \apj, 617, 559

\bibitem[{{Pascucci} {et~al.}(2003){Pascucci}, {Apai}, {Henning}, \&
  {Dullemond}}]{2003ApJ...590L.111P}
{Pascucci}, I., {Apai}, D., {Henning}, T., \& {Dullemond}, C.~P. 2003, \apjl,
  590, L111

\bibitem[{{Rebolo} {et~al.}(1995){Rebolo}, {Zapatero-Osorio}, \&
  {Martin}}]{1995Natur.377..129R}
{Rebolo}, R., {Zapatero-Osorio}, M.~R., \& {Martin}, E.~L. 1995, \nat, 377, 129

\bibitem[{{Reipurth} \& {Clarke}(2001)}]{2001AJ....122..432R}
{Reipurth}, B. \& {Clarke}, C. 2001, \aj, 122, 432

\bibitem[{{Reipurth} {et~al.}(2000){Reipurth}, {Yu}, {Heathcote}, {Bally}, \&
  {Rodr{\'{\i}}guez}}]{2000AJ....120.1449R}
{Reipurth}, B., {Yu}, K., {Heathcote}, S., {Bally}, J., \& {Rodr{\'{\i}}guez},
  L.~F. 2000, \aj, 120, 1449

\bibitem[{{Saito} {et~al.}(2002){Saito}, {Aikawa}, {Herbst}, {Ohishi},
  {Hirota}, {Yamamoto}, \& {Kaifu}}]{2002ApJ...569..836S}
{Saito}, S., {Aikawa}, Y., {Herbst}, E., {et~al.} 2002, \apj, 569, 836

\bibitem[{{Sterzik} \& {Durisen}(1998)}]{1998A&A...339...95S}
{Sterzik}, M.~F. \& {Durisen}, R.~H. 1998, \aap, 339, 95

\bibitem[{{Sterzik} {et~al.}(2004){Sterzik}, {Pascucci}, {Apai}, {van der
  Bliek}, \& {Dullemond}}]{2004A&A...427..245S}
{Sterzik}, M.~F., {Pascucci}, I., {Apai}, D., {van der Bliek}, N., \&
  {Dullemond}, C.~P. 2004, \aap, 427, 245

\bibitem[{{Umbreit} {et~al.}(2004){Umbreit}, {Burkert}, {Henning}, {Mikkola},
  \& {Spurzem}}]{Umbreit}
{Umbreit}, S., {Burkert}, A., {Henning}, T., {Mikkola}, S., \& {Spurzem}, R.
  2004, \apj, in press

\bibitem[{{Wainscoat} {et~al.}(1992){Wainscoat}, {Cohen}, {Volk}, {Walker}, \&
  {Schwartz}}]{1992ApJS...83..111W}
{Wainscoat}, R.~J., {Cohen}, M., {Volk}, K., {Walker}, H.~J., \& {Schwartz},
  D.~E. 1992, \apjs, 83, 111

\bibitem[{{White} {et~al.}(1999){White}, {Ghez}, {Reid}, \&
  {Schultz}}]{1999ApJ...520..811W}
{White}, R.~J., {Ghez}, A.~M., {Reid}, I.~N., \& {Schultz}, G. 1999, \apj, 520,
  811

\bibitem[{{Whittet} {et~al.}(2001){Whittet}, {Gerakines}, {Hough}, \&
  {Shenoy}}]{2001ApJ...547..872W}
{Whittet}, D.~C.~B., {Gerakines}, P.~A., {Hough}, J.~H., \& {Shenoy}, S.~S.
  2001, \apj, 547, 872

\bibitem[{{Whitworth} \& {Zinnecker}(2004)}]{2004A&A...427..299W}
{Whitworth}, A.~P. \& {Zinnecker}, H. 2004, \aap, 427, 299

\bibitem[{{Young} {et~al.}(2003){Young}, {Shirley}, {Evans}, \&
  {Rawlings}}]{2003ApJS..145..111Y}
{Young}, C.~H., {Shirley}, Y.~L., {Evans}, N.~J., \& {Rawlings}, J.~M.~C. 2003,
  \apjs, 145, 111

\end{thebibliography}
\end{document}